# Longitudinal retinal imaging and auxiliary tasks for discrete-time survival prediction of referable diabetic retinopathy and maculopathy: a retrospective cohort study

Paul Nderitu FRCOphth[1]*, Laura Webster[2], Samantha S Mann FRCOphth[2,3], Christos Bergeles PhD[4], Timothy L Jackson PhD FRCOphth[1,5]

[1]Section of Ophthalmology, Faculty of Life Sciences and Medicine, King's College London, London, UK

[2]South East London Diabetic Eye Screening Programme, Guy's and St Thomas' NHS Foundation Trust, London, UK

[3]Department of Ophthalmology, Guy's and St Thomas' NHS Foundation Trust, London, UK

[4]School of Biomedical Engineering & Imaging Sciences, King's College London, London, UK

[5]King's Ophthalmology Research Unit, King's College Hospital, London, UK

*Correspondence to: Dr Paul Nderitu, Section of Ophthalmology, King's College London, London, UK; p.nderitu@doctors.org.uk

## Abstract

**Background** Patients with referable diabetic retinopathy (rDR) or referable maculopathy (rM) in screening are referred to the hospital eye service for monitoring and treatment. Deep learning can predict progression to incident rDR/rM, but existing models use single visit data and ignore longitudinal images available in clinical practice. We assessed whether multi-visit (MV) images, and auxiliary objectives, improve prognostic predictions.

**Methods** In this retrospective cohort study of macula-centred photographs from the South-East London Diabetic Eye Screening Programme (UK), eyes without referable disease at baseline with a follow-up at 20–28 months were split into development (72 559 eyes, 36 648 patients) and test (17 917 eyes, 9 071 patients) sets. A time-aware discrete-time survival model predicted incident rDR (R2 or R3), rM (M1), or either (rDR/rM). We compared baseline-only input with up to three visits (MV), alone or auxiliary objectives of future-visit age (MV+age) or future-visit latent prediction (MV+latent). Models were evaluated using the 28-month cumulative risk AUROC and paired DeLong tests with Benjamini–Hochberg correction.

**Findings** In the test set, 308 eyes (1·7%) developed rDR/rM, 58 (0·3%) rDR, and 275 (1·5%) rM. MV images increased rDR/rM AUROC from 0·749 to 0·786 (+0·037, 95% CI 0·012–0·061; q=0·011) and rM AUROC from 0·749 to 0·788 (+0·039, 0·013–0·065; q=0·010), both significant after correction for multiple testing. Auxiliary objectives did not improve MV model performance. In eyes with two or three visits, prior visits accounted for 30–38% of positive Grad-CAM attributions.

**Interpretation** Prior screening images improved prediction of incident maculopathy and the composite rDR/rM outcome, with gains driven by longitudinal images, not auxiliary supervision. External validation and prospective studies are required before the deployment of models to individualise screening.

## Introduction

An estimated 537 million adults were living with diabetes in 2021, a number projected to reach 783 million by 2045.[1] Diabetic retinopathy affects about one in five people with diabetes and remains a leading cause of preventable vision loss in working-age adults.[2,3] The English NHS Diabetic Eye Screening Programme (DESP) photographs more than 3.2 million people each year and refers those with referable diabetic retinopathy (rDR) or referable maculopathy (rM) to hospital eye services.[4] Most attendees never develop referable disease, and screening of low-risk people consumes capacity that could be redirected to those at higher risk.[5] Randomised and observational evidence shows that individualised, risk-based screening intervals are safe and could reduce appointments and costs, but existing risk engines depend on systemic data such as glycated haemoglobin that are not routinely available at the time of screening.[6,7]

Deep learning detects referable diabetic retinopathy in fundus photographs with high accuracy,[8,9] and prognostic deep learning systems can predict future progression from images alone.[10–16] These systems predict progression to referable or vision-threatening disease 1–5 years ahead, with AUROCs of about 0·7–0·9.[10–15] We previously showed that a single screening visit of two-field photographs predicts incident rDR and rM at 1, 2 and 3 years, with image-based models outperforming models using clinical and demographic data alone.[15] However, prior prognostic image-based systems for predicting referable DR used data from a single baseline visit.[10–15] People attending DR screening typically have several years of archived photographs of the same eye. Change over time, such as accumulating microaneurysms or exudates approaching the fovea, are plausible prognostic signals for disease progression that single visit images do not capture. Single visit and sequential imaging models for age-related macular degeneration prediction have been described,[17,18] but the use of routine longitudinal fundus images to improve the prediction of incident referable diabetic eye disease is largely unexplored.

A second question is whether auxiliary objectives could improve sequential prognostic models. Auxiliary tasks can regularise networks trained on rare outcomes by providing a denser training signal.[19] Two candidate objectives follow directly from the longitudinal setting. First, age can be predicted accurately from fundus photographs, and the gap between retinal and chronological age carries prognostic information,[20,21] so predicting age at the future visit might encourage the model to encode ageing-related retinal changes which are potential risk-factors of disease progression. Second, self-supervised methods that predict the representation of an unseen view from an observed one could assist models to extract useful prognostic features.[22,23]

We trained and tested a time-aware multi-visit discrete-time survival model for incident rDR, rM and either outcome (rDR/rM) over a 20–28-month horizon on macula-centred photographs from a large and ethnically diverse UK screening programme. Our primary objective was to compare baseline-only with multi-visit prediction. Secondary objectives were to assess whether auxiliary future-visit age supervision or latent prediction improved performance, and to quantify which visits contributed to model predictions.

## Methods

### Study design and participants

This retrospective cohort study used data from the South-East London Diabetic Eye Screening Programme (SEL-DESP), which offers annual screening to people with diabetes aged 12 years or older. At each screen, two 45° colour fundus photographs per eye (macula-centred and optic disc-centred) are acquired after mydriasis and graded by accredited graders according to NHS DESP grading definitions, with arbitration of discordant grades.[24] We included screens performed between September 2013 and December 2019. Images were curated for laterality, field and gradability using a previously validated deep learning pipeline,[25] and only gradable macula-centred photographs were used. The study is reported in accordance with the TRIPOD+AI statement.[26]

An eye was eligible if it had a baseline screen without referable disease (retinopathy grade R0 or R1 and maculopathy grade M0) and a graded follow-up screen 20–28 months after baseline. Each eligible eye contributed one baseline visit and up to two additional gradable screens prior to the baseline for multi-visit models. Eyes with no earlier screen contributed the baseline image only.

Eyes were assigned at patient level to a development set (72 559 eyes, 36 648 patients, 80%) and a test set (17 917 eyes, 9 071 patients, 20%), with no patient contributing to more than one set (figure 1A). Within the development set, one-eighth of patients were then assigned to a tuning set using stratified group k-fold partitioning (eight folds, grouped by patient and stratified by the rDR/rM outcome) with the remaining seven-eighths formed the training set (figure 1A). Health Research Authority approval and a favourable opinion from the East Midlands Leicester South Research Ethics Committee (REC, 20/EM/0250) were obtained prior to study commencement.

### Outcomes

The outcomes were incident rDR, defined as pre-proliferative (R2) or proliferative (R3) retinopathy; incident rM, defined as referable maculopathy (M1); and either outcome (rDR/rM). Outcomes were ascertained by screening programme graders at the first follow-up screen within the 20–28-month window (figure 1B). The window accommodates the variation in timing of routine recall around the nominal 24-month interval and was divided into K=2 equal discrete intervals (20–24 and 24–28 months). Because outcomes are observed only at screening visits, onset is interval-censored; each eye was assigned to the interval containing its follow-up screen.

### Discrete-time survival formulation

For each outcome, the model outputs a hazard $\lambda_k$ for each interval k, and the cumulative risk of the outcome by the end of interval k is $F(k)=1-\prod_{j\leq k}(1-\lambda_j)$.[27] Training targets were per-interval binary outcomes and were considered positive from the interval containing an event onward. For eyes without an event whose follow-up screen fell in the first interval, the second interval was unobserved hence was masked from the loss computation. Eyes without an event by the final outcome interval (24–28 months) were treated as fully observed negatives. The cumulative risk at 28 months, was used as the evaluation risk score.

### Model architecture

The model architecture is summarised in figure 1C. Each input photograph was centre-cropped to a square, resized to 448×448 pixels and normalised with ImageNet statistics. A shared ConvNeXt-V2-Large encoder,[28] initialised with ImageNet-pretrained weights, mapped each visit image to a 1536-dimensional embedding by global average pooling. Because screening intervals are irregular, the time of each visit relative to baseline (in months) was added to its embedding using a continuous-time sinusoidal encoding, adapted from transformer positional encodings[29] with a base frequency chosen to resolve the 0–48-month clinical range. The time-encoded sequence was processed by a pre-norm transformer encoder layer (eight attention heads, feed-forward width 3 072) with a key-padding mask so that absent visits did not contribute. The transformer output at the baseline visit, which attends to the earlier visits through self-attention, was taken as the single summary latent h; for eyes with only a baseline visit, h depends on that visit alone. Three survival heads operated on the summary latent *h* with linear heads predicting the interval hazard logits for rDR and rM with dropout. A small multilayer perceptron combined these two sets of logits into the combined rDR/rM outcome to predict the hazard. The baseline-only configuration used the identical architecture with a single input visit. For multi-visit models, eyes with fewer than three visits were zero-padded, and padded positions were excluded by the attention mask.

### Auxiliary objectives

A future-state predictor made up of a residual multilayer perceptron with layer normalisation and SiLU activations, took *h* plus a sinusoidal encoding of the time from baseline to the follow-up screen (Δt) to produce a predicted future-visit embedding $\hat{z}$. Two auxiliary objectives used $\hat{z}$. For latent prediction, the target $z$ was the embedding of the actual follow-up image computed by an exponential moving average (EMA) copy of the network's image encoder (momentum 0·99 per optimiser step). Following bootstrap and joint-embedding predictive methods,[22,23] the EMA target received no gradient. The alignment loss was a weighted (each 1.0) sum of cosine and smooth-L1 loss between $\hat{z}$ and $z$. For future-age supervision, a multilayer perceptron mapped $\hat{z}$ to the patients age at the follow-up visit. The follow-up image latent and age were used only during training for supervision. At inference, survival predictions depend only on the visit history.

**Model configurations and training**

We compared four configurations that differed only in input visits and auxiliary objectives (figure 1D). The baseline-only configuration (BL) used the baseline visit alone with the survival objective. The multi-visit configuration (MV) used up to three visits with the survival objective. MV+age added future-age supervision, and MV+latent added future-visit latent prediction. The survival loss was the mean across the three outcomes of a censoring-masked binary focal loss ($\gamma$=2; $\alpha$=0·20 for rDR/rM and rM, and 0·60 for rDR).[30] The total loss was $L=\lambda_{surv}L_{surv}+\lambda_{age}L_{age}+\lambda_{lat}L_{lat}$, with $\lambda_{surv}$= 5.0 and $\lambda_{age}$=0.5 and $\lambda_{lat}$=0.5 when the corresponding auxiliary objective was active, otherwise the loss weighting was set to zero.

To counter the rarity of rDR, training eyes were drawn by weighted random sampling so that eyes with incident rDR made up 12% of sampled eyes. Augmentation comprised random vertical flipping (p=0·5), colour jitter (brightness 0·15, contrast 0·5, saturation 0·25, hue 0·1) and Gaussian blur (p=0·1). Models were optimised with AdamW[31] using a base learning rate of 1e-5 for the encoder and a ten-fold higher rate for the temporal encoder and all heads, with weight decay 3e-4, cosine annealing to $1\times10^{-6}$, mixed-precision arithmetic, a batch size of 4 with 8 gradient-accumulation steps, and distributed data-parallel training on two graphics processing units. Training ran for up to 15 epochs with early stopping (patience five epochs) on the validation rDR/rM Harrell's concordance index, and the checkpoint with the highest validation concordance index was retained.

## Figure 1: Study design

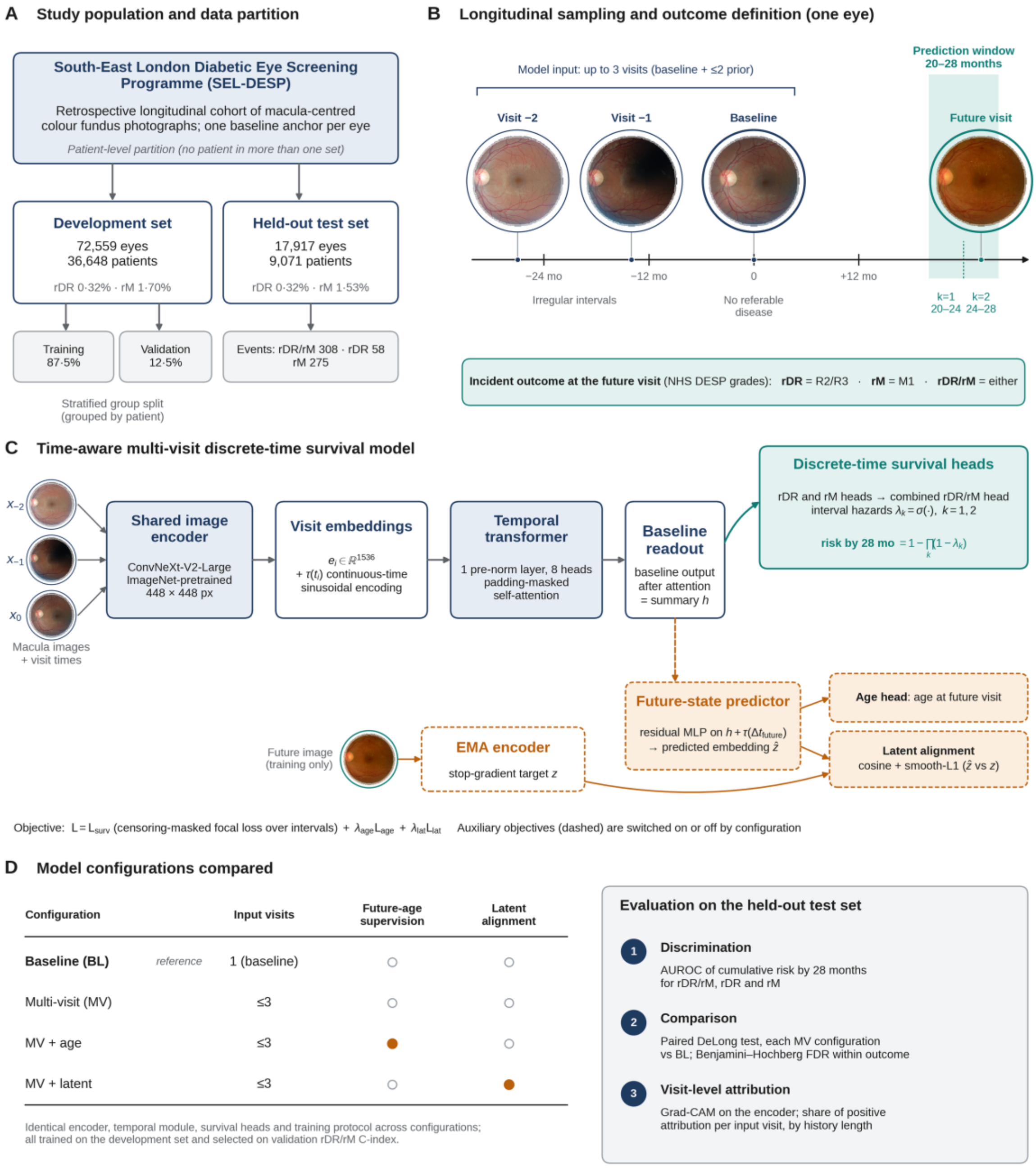

(A) Patient-level partition of the South-East London Diabetic Eye Screening Programme cohort into development (training and validation) and test sets. (B) Longitudinal sampling for one eye. Up to two prior screens and the baseline screen form the model input, and the outcome is graded at the follow-up screen within the 20–28-month prediction window, which is divided into two discrete intervals. Images show a test-set eye that developed referable retinopathy and maculopathy. (C) Model architecture. A shared ConvNeXt-V2-Large encoder embeds each visit, a continuous-time encoding represents irregular visit timing, and a padding-masked temporal transformer summarises the history, which is read out at the baseline visit. Discrete-time survival heads output interval hazards for referable retinopathy (rDR), referable maculopathy (rM) and either outcome (rDR/rM). Optional auxiliary branches (dashed) predict the future-visit latent from the history and time to follow-up. This prediction is either aligned with the embedding of the observed follow-up image from an exponential moving average (EMA) encoder or used to predict age at follow-up. (D) Model configurations compared and evaluation.

### Visit-level attribution

To assess how much the MV model relied on each input visit, we computed Grad-CAM attributions[34] with Captum.[35] Attribution was taken with respect to the final encoder stage for the 28-month cumulative risk of each outcome. For each outcome, we randomly sampled 50 test-set eyes with that outcome, using a fixed random seed. Attribution maps were restricted to positive values, which indicate evidence that increased predicted risk, and upsampled to image resolution. For each eye, the share of positive attribution assigned to each visit was that visit's mean positive attribution divided by the sum across the eye's input visits. Because eyes have different numbers of visits, results were stratified by history length (one, two or three visits), indexed by position relative to baseline, and summarised as means with 95% CIs. Maps were displayed on a colour scale normalised across all visits of the same eye, so intensities are comparable between visits.

### Statistical analysis

The primary performance measure was the AUROC of the 28-month cumulative risk for each outcome in the test set. Each multi-visit configuration was compared with BL on the same eyes using the paired DeLong test,[32] which also gave 95% CIs for the differences in AUROC. Within each outcome, p values for the three comparisons with BL were adjusted using the Benjamini–Hochberg procedure,[33] and comparisons with $q<0{\cdot}05$ were considered significant. Direct comparisons between multi-visit configurations were not prespecified. Analyses were done at the eye level and the correlation between the two eyes of a patient was not modelled. Cohort characteristics are summarised descriptively. Analyses used Python 3.8.20, Torch 2.3.1+cu118, PyTorch Lightning 2.3.0, timm 1.0.27, Captum 0.7.0, scikit-learn 1.3.0 and scikit-survival 0.22.2.

## Results

The development set comprised 72 559 eyes of 36 648 patients and the test set 17 917 eyes of 9071 patients (figure 1A; table 1). The two sets were similar: mean age was 61 years in the development set and 62 years in the test set, 54·0% and 54·5% of participants were male, 53·1% and 52·9% were White, and 94·5% and 94·7% had type 2 diabetes, with a mean diabetes duration of 8 years in both. Eyes had a median of four recorded screening visits in the development set and three in the test set. Incidence over the 20–28-month window was low and similar between sets. In the test set, 308 eyes (1·72%) developed rDR/rM, 58 (0·32%) rDR and 275 (1·53%) rM. The corresponding development-set incidences were 1 351(1·86%) for rDR/rM, 232 (0·32%) for rDR and 1 235 (1·70%) for rM.

**Table 1: Characteristics of the development and test sets**

| | Development set | Test set |
|---|---|---|
| Eyes | 72 559 | 17 917 |
| Patients | 36 648 | 9071 |
| Age, years | 61 (14) | 62 (14) |
| Male sex | 54·0% | 54·5% |
| White ethnicity | 53·0% | 52·9% |
| Type 2 diabetes | 94·5% | 94·7% |
| Diabetes duration, years | 8 (7) | 8 (7) |
| Screening visits per eye, median | 4 | 3 |
| **Incident outcome within 20–28 months** | | |
| rDR/rM | 1 351 (1·86%) | 308 (1·72%) |
| rDR (R2 or R3) | 232 (0·32%) | 58 (0·32%) |
| rM (M1) | 1 235 (1·70%) | 275 (1·53%) |

Data are n, mean (SD), %, or n (%). Sex, ethnicity and diabetes type are reported at the patient level. rDR=referable diabetic retinopathy. rM=referable maculopathy.

AUROCs and differences from BL are shown in table 2 and figure 2. BL achieved AUROCs of 0·749 for rDR/rM, 0·812 for rDR and 0·749 for rM. Adding prior visits (MV) increased the AUROC for rDR/rM to 0·786 (difference 0·037, 95% CI 0·012 to 0·061; p=0·0038), and this difference remained significant after correction for multiple testing (q=0·011). The improvement was driven by rM, for which the AUROC increased from 0·749 to 0·788 (0·039, 0·013 to 0·065; p=0·0034; q=0·010). For rDR, MV did not differ significantly from BL (0·820; 0·008, −0·051 to 0·066; p=0·80; q=0·80), and the wide CI reflected the small number of events.

Neither auxiliary objective improved discrimination for rDR/rM or rM beyond that of MV. With future-age supervision, AUROCs were 0·775 for rDR/rM (0·026, 0·002 to 0·051; p=0·037; q=0·056), 0·827 for rDR (0·015, −0·039 to 0·070; p=0·59; q=0·80) and 0·781 for rM (0·032, 0·007 to 0·057; p=0·012; q=0·019); the rM improvement remained significant after correction for multiple testing. With latent prediction, no difference from BL was significant: 0·756 for rDR/rM (0·006, −0·022 to 0·035; p=0·66; q=0·66), 0·777 for rDR (−0·035, −0·101 to 0·031; p=0·30; q=0·80) and 0·767 for rM (0·018, −0·012 to 0·048; p=0·25; q=0·25). For rDR/rM and rM, point estimates for both auxiliary configurations were lower than for MV. For rDR, MV+age had the highest point estimate (0·827) but did not differ significantly from BL.

## Table 2: Discrimination of each configuration for incident referable disease in the test set (n=17 917 eyes)

| Configuration | AUROC | Difference *vs* BL (95% CI) | p value | q value |
|---|---|---|---|---|
| **rDR/rM (308 events)** | | | | |
| BL (reference) | 0·749 | ·· | ·· | ·· |
| MV | 0·786 | 0·037 (0·012 to 0·061) | 0·0038 | **0·011** |
| MV+age | 0·775 | 0·026 (0·002 to 0·051) | 0·037 | 0·056 |
| MV+latent | 0·756 | 0·006 (−0·022 to 0·035) | 0·66 | 0·66 |
| **rDR (58 events)** | | | | |
| BL (reference) | 0·812 | ·· | ·· | ·· |
| MV | 0·820 | 0·008 (−0·051 to 0·066) | 0·80 | 0·80 |
| MV+age | 0·827 | 0·015 (−0·039 to 0·070) | 0·59 | 0·80 |
| MV+latent | 0·777 | −0·035 (−0·101 to 0·031) | 0·30 | 0·80 |
| **rM (275 events)** | | | | |
| BL (reference) | 0·749 | ·· | ·· | ·· |
| MV | 0·788 | 0·039 (0·013 to 0·065) | 0·0034 | **0·010** |
| MV+age | 0·781 | 0·032 (0·007 to 0·057) | 0·012 | **0·019** |
| MV+latent | 0·767 | 0·018 (−0·012 to 0·048) | 0·25 | 0·25 |

AUROC of the 28-month cumulative risk. Differences, 95% CIs and p values are from paired DeLong tests against BL. q values are Benjamini–Hochberg-adjusted within each outcome across the three comparisons; bold indicates q<0·05. AUROC=area under the receiver operating characteristic curve. BL=baseline visit only. MV=multi-visit (up to three visits). MV+age=MV with future-visit age supervision. MV+latent=MV with future-visit latent embedding alignment. rDR=referable diabetic retinopathy. rM=referable maculopathy.

## Figure 2: Difference in AUROC of each multi-visit configuration compared with the baseline-only model

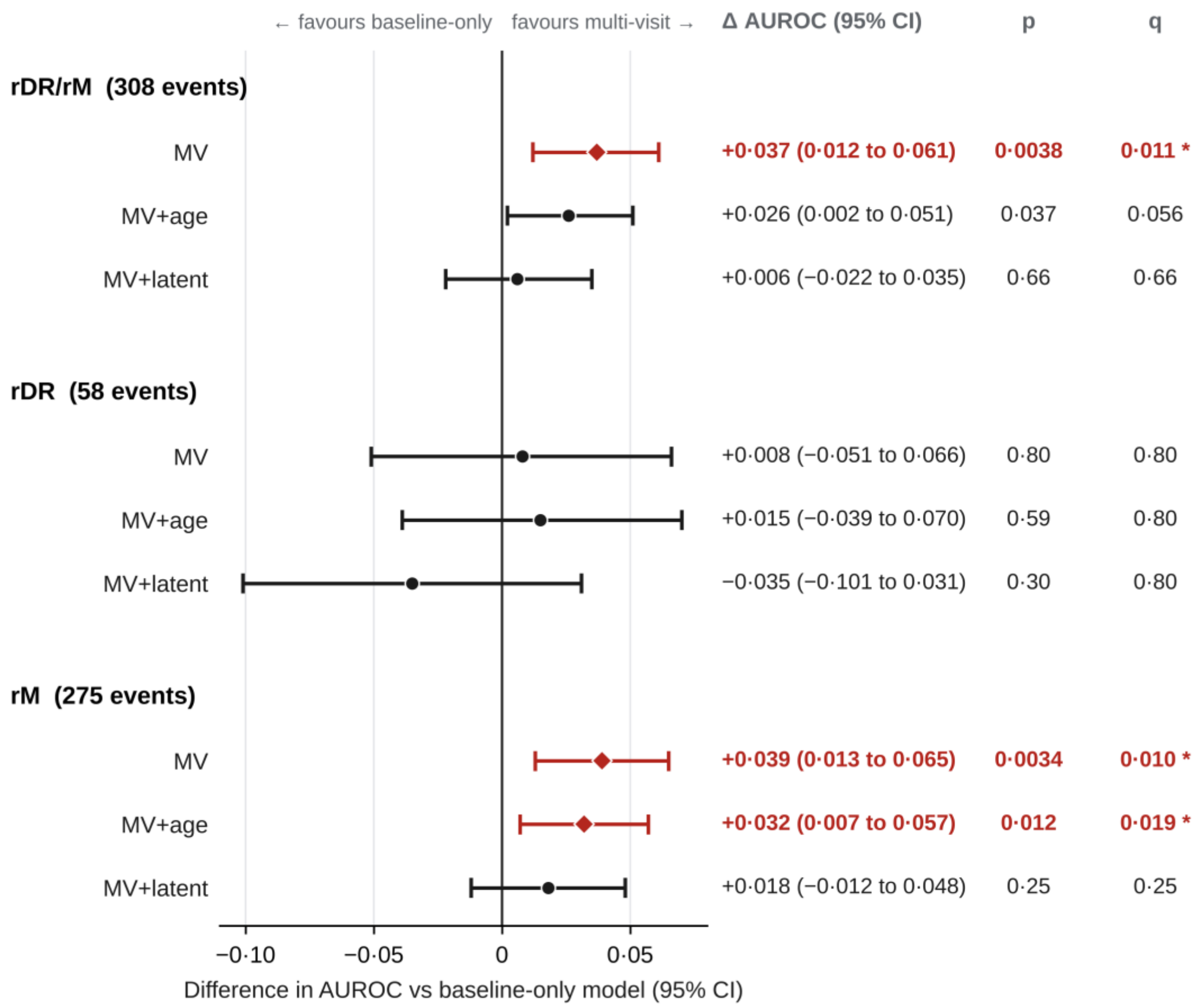

* Benjamini–Hochberg q<0·05 (adjusted within each outcome). Red diamonds mark these comparisons.

Points show the difference in AUROC (comparator minus BL) for the 28-month cumulative risk in the test set, and bars show 95% CIs from paired DeLong tests. Annotations give the difference with 95% CI, p value and Benjamini–Hochberg q value (adjusted within each outcome). Highlighted comparisons (red) had q<0·05. AUROC=area under the receiver operating characteristic curve. BL=baseline-only model. rDR=referable diabetic retinopathy. rM=referable maculopathy.

Visit-level attribution is summarised in figure 3. In eyes with two input visits, the baseline visit accounted for a mean of 62–70% of positive attribution across the three outcomes and the prior visit for 30–38%. In eyes with three visits, the baseline visit accounted for 64–67%, the visit before baseline (median 14 months earlier) for 18–20%, and the earliest visit (median 27 months earlier) for 15–16%. In every stratum, the combined share of the prior visits (means 30–38%) was below the share expected if all visits were weighted equally (50% with two visits and 67% with three), but no eye relied on the baseline visit alone (minimum prior-visit share 14%). Attribution therefore declined with distance from baseline, but prior visits retained a substantial share. Representative maps (figure 4) show focal attribution at baseline in the macula and around the optic disc, with lower-intensity foci at earlier visits, including at the vascular arcades and near the optic disc.

**Figure 3: Share of positive attribution assigned to each input visit, stratified by history length**

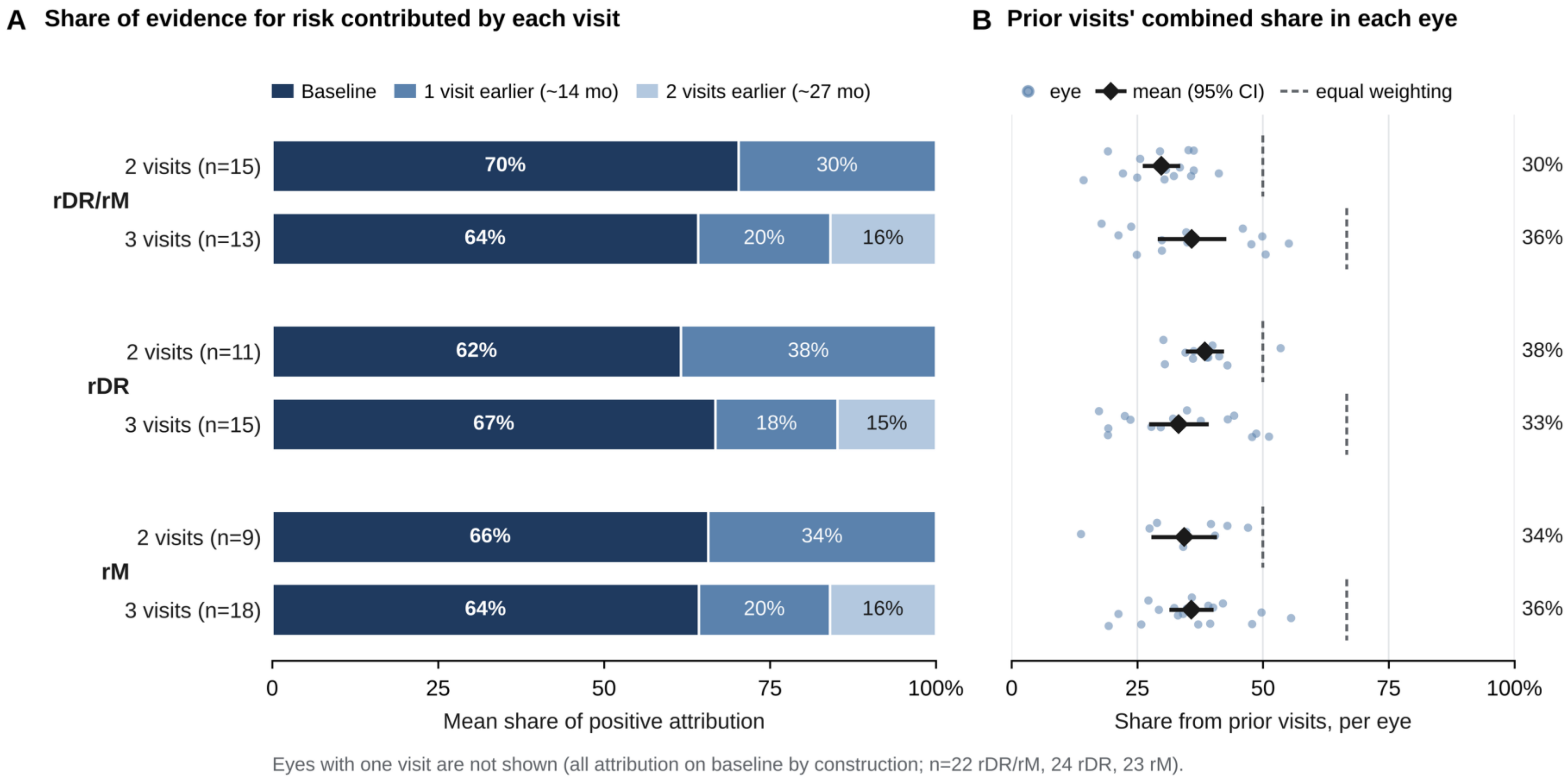

Grad-CAM attribution of the 28-month cumulative risk from the multi-visit model, in test-set eyes that developed each outcome (50 eyes per outcome), stratified by the number of input visits. (A) Mean share of each eye's positive attribution assigned to the baseline visit and to the visits one (median 14 months) and two (median 27 months) before baseline; shares sum to 100% within each row. (B) Combined share of the prior visits in each eye (points), with mean and 95% CI; dashed lines show the share expected if all visits were weighted equally (50% with two visits, 67% with three). Eyes with a single visit (22 rDR/rM, 24 rDR and 23 rM) are not shown because all attribution falls on the baseline visit by construction. n is the number of eyes in each stratum. rDR=referable diabetic retinopathy. rM=referable maculopathy.

### Figure 4: Visit-level Grad-CAM attribution maps from the multi-visit model

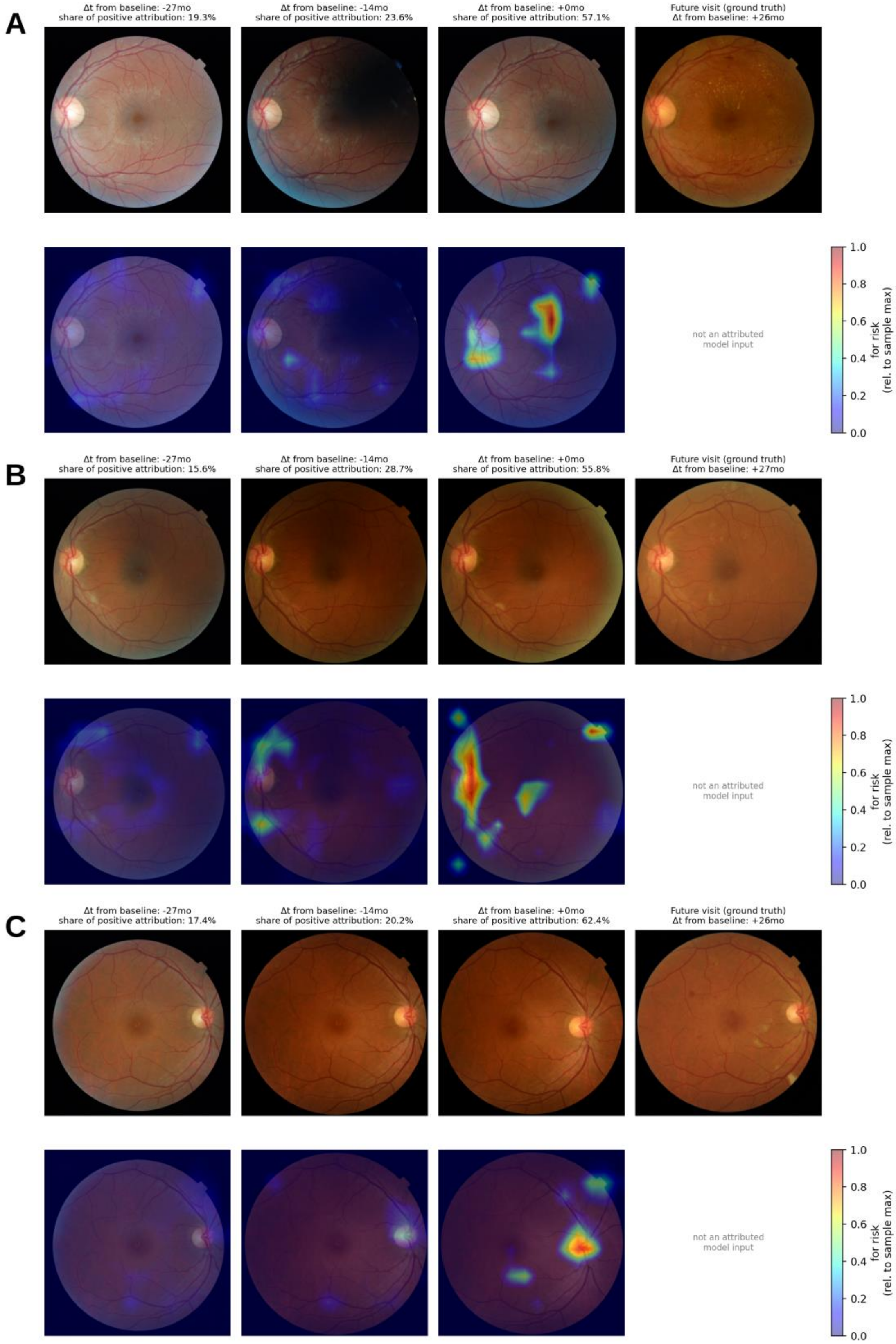

(A–C) Three test-set eyes that developed rDR with three visits. The target was the 28-month cumulative rDR risk. For each eye, the top row shows the input photographs, with time from baseline and the visit's share of positive attribution. The follow-up photograph was not use as an input. The bottom row shows positive attributions per input, normalised across visits so intensities are comparable between visits.

## Discussion

In this large screening cohort, adding up to two prior screening photographs to a deep learning survival model improved prediction of incident referable disease over roughly 2 years, and the gain was largely driven by incident referable maculopathy. Improvement for the combined rDR/rM outcome also remained significant after correction for multiple testing. We found no evidence that either auxiliary objective, predicting future-visit age or latent, added benefit beyond the longitudinal images. Attribution analysis showed that the multi-visit model used features from earlier visits rather than relying on the most recent image alone. Together, these findings suggest that the prognostic benefit in predicting incident disease comes from the longitudinal images rather than from auxiliary supervision of future age or future latent prediction.

Earlier prognostic deep learning systems for diabetic retinopathy compute image risk from a single visit.[7,10–16] Our results show that historical images carry additional prognostic information, notably for maculopathy. A plausible explanation is that maculopathy in screening is defined by exudates and microaneurysms or haemorrhages within specified distances of the fovea.[24] Progression towards this threshold is a spatial and temporal process, so the trajectory of lesions across visits may provide better estimate of future progressions compared to a single time point image. Maculopathy was also about five times more frequent than retinopathy in our cohort, which gave the comparison more power. With only 58 rDR events in the test set, the study could not so easily establish whether longitudinal input helps for retinopathy. BL already discriminated rDR relatively well (AUROC 0·812), which may leave less room for improvement.

The auxiliary objectives were motivated by evidence that fundus photographs encode age[20,21] and that predicting representations of unseen views is a powerful learning signal.[22,23] Neither translated into better discrimination than multi-visit input alone; future-age supervision still improved maculopathy prediction relative to BL, but with a smaller point estimate than MV alone. Predicting future age may not provide significant prognostic information for retinopathy and maculopathy progression that the model has not already derived from macula images alone. Predicting the future latent requires that the model predict the entire embedding of the follow-up image which is a challenging task. That embedding also reflects acquisition factors that are unpredictable from the history, such as camera, illumination, focus, pupil size and media opacity, as well as any disease change. When events are rare, an objective dominated by this variance can compete with the survival objective for representational capacity, and it may destabilise the shared encoder through the EMA target. Targets restricted to disease-relevant features, such as lesion segmentations or future grades, or applying auxiliary objectives during pretraining rather than fine-tuning, may be more useful and merit testing.

The visit-level attributions provide a clue as to whether the network's evidence for high risk came from the most recent image only. In eyes with two or three visits, about a third of positive attribution (30–38%) came from earlier visits, although the share declined with distance from baseline. Because the summary is read out at the baseline visit, which attends to the earlier visits, the baseline image anchors each prediction, consistent with its larger share. Therefore, whilst recent images are more informative, history visit images still carry additional prognostic value which is consistent with the discrimination results.

These findings have practical implications. Patients attend diabetic retinopathy screening programmes every one to two years. Therefore, most patients have readily available multi-visit model data. This would also be the case in routine hospital eye care where chronic retinal conditions mean patients require review over multiple visits. Foundation models pretrained on large retinal image datasets[37] could therefore serve as encoders within a longitudinal framework and may improve data efficiency for longitudinal modelling for multiple retinal and Oculomics use cases.

This study has limitations. First, whilst the data is derived from a large, ethnically diverse screening programme, the model has not been externally validated. Second, rDR events were rare, which limited precision for that outcome. Third, we used only the macula-centred field; the disc-centred field may add information, particularly changes in the nasal periphery. Fourth, eye-level analyses treated fellow eyes as independent, which may make CIs optimistically narrower, however, clinical decisions in DR screening are usually made at the eye level. Fifth, eyes with longer histories may differ systematically from those with shorter histories, for example in diabetes duration or attendance, and we did not assess performance by history length. Sixth, we evaluated a single horizon and prognostic gains with historical visit data may differ with different prediction horizons. Seventh, reference grades came from routine screening and are subject to grader variability, although arbitration and quality assurance in the NHS DESP mitigate this. Finally, attribution analyses have known limitations with Grad-CAM only providing coarse maps and findings reflecting the model attention rather than causal effects. Attributions were also computed on a sample of positive eyes and attributions in negative cases require future study.

In conclusion, prior screening images improved prediction of incident referable maculopathy and the composite rDR/rM outcome, and longitudinal data rather than auxiliary supervision drove the performance gain. Prognostic models in diabetic eye screening, and potentially in other retinal and systemic diseases monitored by repeated imaging, should make use of the image histories that health systems already hold. External validation and prospective model evaluation are the next steps towards realising individualised screening intervals using multi-visit data.

### Contributors

PN conceived the study, curated data, developed the models, did the analysis and wrote the first draft; LW and SSM provided data access and clinical expertise; CB and TLJ supervised the study. All authors reviewed the manuscript. PN accessed and verified the underlying data, and that all authors had final responsibility for the decision to submit.

**Funding** Diabetes UK (20/0006144) and Wellcome/EPSRC Centre for Medical Engineering grant (WT203148/Z/16/Z)

### Role of the funding source

Study funders did not have access to the study data, nor did they influence the study design, DLS development, data analysis, or manuscript preparation. The corresponding author had the final responsibility for the decision to submit the manuscript.

### Declaration of interests

L Webster and C Bergeles have no conflicts of interest to declare. P Nderitu has received speaker fees from Boehringer Ingelheim, Roche, Alimera and is an advisor with equity for Datta Yakka, SS Mann has received speaker and advisory board fees from Bayer and Allergan and has received research grants from Novartis. T Jackson is a consultant or adviser to 2CTech, Alcon, Annexon, Dutch Ophthalmic Research Centre, iLumen, Opthea, Outlook Therapeutics, and Oxurion. He had free use of equipment for a non-commercial clinical trial from LKC. He has also received payment as a clinical expert by solicitors acting for REGENERON. T Jacksons' employer (King's College Hospital) receives funding for participants enrolled on commercial clinical trials of diabetic retinopathy including THR149-002 (OXURION), NEON NPDR (BAYER), RHONE-X (ROCHE) and ALTIMETER (ROCHE).

### Code availability

Open-source resources and code were used to develop and validate the prognostic deep learning systems including python, pytorch, timm, Captum libraries.

### Data sharing

The screening data are not publicly available owing to information-governance restrictions.

### Acknowledgments

The model development study was funded by Diabetes UK via a research training fellowship grant to Dr Paul Nderitu (20/0006144). The study was also supported by a Wellcome EPSRC Centre for Medical Engineering grant (WT 203148/Z/16/Z) to Dr Christos Bergeles.

## TRIPOD+AI checklist

| Section/Topic | Item | Dev/Eval* | Checklist item | Reported on page |
|---|---|---|---|---|
| **TITLE** | | | | |
| **Title** | 1 | D;E | Identify the study as developing or evaluating the performance of a multivariable prediction model, the target population, and the outcome to be predicted | p 1 (Title) |
| **ABSTRACT** | | | | |
| **Abstract** | 2 | D;E | See TRIPOD+AI for Abstracts checklist | p 2 (Summary) |
| **INTRODUCTION** | | | | |
| **Background** | 3a | D;E | Explain the healthcare context (including whether diagnostic or prognostic) and rationale for developing or evaluating the prediction model, including references to existing models | p 4 (Introduction) |
| | 3b | D;E | Describe the target population and the intended purpose of the prediction model in the context of the care pathway, including its intended users (e.g., healthcare professionals, patients, public) | p 4 (Introduction)<br>p 54 (Methods: Study design and participants) |
| | 3c | D;E | Describe any known health inequalities between sociodemographic groups | *Not reported* |
| **Objectives** | 4 | D;E | Specify the study objectives, including whether the study describes the development or validation of a prediction model (or both) | p 5 (Introduction) |
| **METHODS** | | | | |
| **Data** | 5a | D;E | Describe the sources of data separately for the development and evaluation datasets (e.g., randomised trial, cohort, routine care or registry data), the rationale for using these data, and representativeness of the data | p 5 (Methods: Study design and participants)<br>p 5 (Methods: Study design and participants) |
| | 5b | D;E | Specify the dates of the collected participant data, including start and end of participant accrual; and, if applicable, end of follow-up | p 5 (Methods: Study design and participants) |
| **Participants** | 6a | D;E | Specify key elements of the study setting (e.g., primary care, secondary care, general population) including the number and location of centres | p 5 (Methods: Study design and participants)<br>*Number of screening sites not stated* |
| | 6b | D;E | Describe the eligibility criteria for study participants | p 5 (Methods: Study design and participants) |
| | 6c | D;E | Give details of any treatments received, and how they were handled during model development or evaluation, if relevant | *Not applicable (eyes had no referable disease at baseline); not stated explicitly* |
| **Data preparation** | 7 | D;E | Describe any data pre-processing and quality checking, including whether this was similar across relevant sociodemographic groups | p 5 (Methods: Study design and participants)<br>p 6 (Methods: Model architecture) |
| **Outcome** | 8a | D;E | Clearly define the outcome that is being predicted and the time horizon, including how and when assessed, the rationale for choosing this outcome, and whether the method of outcome assessment is consistent across sociodemographic groups | pp 5–6 (Methods: Outcomes)<br>p 6 (Methods: Discrete-time survival formulation) |
| | 8b | D;E | If outcome assessment requires subjective interpretation, describe the qualifications and demographic characteristics of the outcome assessors | p 5 (Methods: Study design and participants) |
| | 8c | D;E | Report any actions to blind assessment of the outcome to be predicted | *Not reported* |
| **Predictors** | 9a | D | Describe the choice of initial predictors (e.g., literature, previous models, all available predictors) and any pre-selection of predictors before model building | p 4 (Introduction)<br>p 5 (Methods: Study design and participants) |

| Section/Topic | Item | Dev/Eval* | Checklist item | Reported on page |
| --- | --- | --- | --- | --- |
| | 9b | D;E | Clearly define all predictors, including how and when they were measured (and any actions to blind assessment of predictors for the outcome and other predictors) | p 6 (Methods: Model architecture) |
| | 9c | D;E | If predictor measurement requires subjective interpretation, describe the qualifications and demographic characteristics of the predictor assessors | *Not applicable (predictors are photographs; no subjective measurement)* |
| **Sample size** | 10 | D;E | Explain how the study size was arrived at (separately for development and evaluation), and justify that the study size was sufficient to answer the research question. Include details of any sample size calculation | *Not reported (all eligible eyes were used; no sample size calculation is described)* |
| **Missing data** | 11 | D;E | Describe how missing data were handled. Provide reasons for omitting any data | p 5 (Methods: Study design and participants)<br>p 6 (Methods: Model architecture) |
| **Analytical methods** | 12a | D | Describe how the data were used (e.g., for development and evaluation of model performance) in the analysis, including whether the data were partitioned, considering any sample size requirements | p 5 (Methods: Study design and participants) |
| | 12b | D | Depending on the type of model, describe how predictors were handled in the analyses (functional form, rescaling, transformation, or any standardisation). | p 6 (Methods: Model architecture) |
| | 12c | D | Specify the type of model, rationale, all model-building steps, including any hyperparameter tuning, and method for internal validation | p 6 (Methods: Model architecture)<br>pp 6–7 (Methods: Auxiliary objectives)<br>p 7 (Methods: Model configurations and training) |
| | 12d | D;E | Describe if and how any heterogeneity in estimates of model parameter values and model performance was handled and quantified across clusters (e.g., hospitals, countries). See TRIPOD-Cluster for additional considerations | *Not applicable (single programme; not examined)* |
| | 12e | D;E | Specify all measures and plots used (and their rationale) to evaluate model performance (e.g., discrimination, calibration, clinical utility) and, if relevant, to compare multiple models | p 8 (Methods: Statistical analysis) |
| | 12f | E | Describe any model updating (e.g., recalibration) arising from the model evaluation, either overall or for particular sociodemographic groups or settings | *Not applicable (no model updating)* |
| | 12g | E | For model evaluation, describe how the model predictions were calculated (e.g., formula, code, object, application programming interface) | p 6 (Methods: Discrete-time survival formulation) |
| **Class imbalance** | 13 | D;E | If class imbalance methods were used, state why and how this was done, and any subsequent methods to recalibrate the model or the model predictions | p 7 (Methods: Model configurations and training)<br>p 7 (Methods: Model configurations and training)<br>*Predictions were not recalibrated* |
| **Fairness** | 14 | D;E | Describe any approaches that were used to address model fairness and their rationale | *Not reported* |
| **Model output** | 15 | D | Specify the output of the prediction model (e.g., probabilities, classification). Provide details and rationale for any classification and how the thresholds were identified | p 6 (Methods: Discrete-time survival formulation) |
| **Training versus evaluation** | 16 | D;E | Identify any differences between the development and evaluation data in healthcare setting, eligibility criteria, outcome, and predictors | p 8 (Results)<br>p 16 (Table 1) |

| Section/Topic | Item | Dev/Eval* | Checklist item | Reported on page |
|---|---|---|---|---|
| **Ethical approval** | 17 | D;E | Name the institutional research board or ethics committee that approved the study and describe the participant-informed consent or the ethics committee waiver of informed consent | p 5 (Methods: Study design and participants) |
| **OPEN SCIENCE** | | | | |
| **Funding** | 18a | D;E | Give the source of funding and the role of the funders for the present study | p 8 (Role of the funding source)<br>p 2 (Funding)<br>p 13 (Acknowledgments) |
| **Conflicts of interest** | 18b | D;E | Declare any conflicts of interest and financial disclosures for all authors | p 12 (Declaration of interests) |
| **Protocol** | 18c | D;E | Indicate where the study protocol can be accessed or state that a protocol was not prepared | *Not reported* |
| **Registration** | 18d | D;E | Provide registration information for the study, including register name and registration number, or state that the study was not registered | *Not reported* |
| **Data sharing** | 18e | D;E | Provide details of the availability of the study data | p 13 (Data sharing) |
| **Code sharing** | 18f | D;E | Provide details of the availability of the analytical code | p 13 (Code availability)<br>*Study code itself not stated as available* |
| **PATIENT & PUBLIC INVOLVEMENT** | | | | |
| **Patient & Public Involvement** | 19 | D;E | Provide details of any patient and public involvement during the design, conduct, reporting, interpretation, or dissemination of the study or state no involvement. | *Not reported* |
| **RESULTS** | | | | |
| **Participants** | 20a | D;E | Describe the flow of participants through the study, including the number of participants with and without the outcome and, if applicable, a summary of the follow-up time. A diagram may be helpful. | p 8 (Results)<br>p 17 (Figure 1) |
| | 20b | D;E | Report the characteristics overall and, where applicable, for each data source or setting, including the key dates, key predictors (including demographics), treatments received, sample size, number of outcome events, follow-up time, and amount of missing data. A table may be helpful. Report any differences across key demographic groups. | p 8 (Results)<br>p 16 (Table 1) |
| | 20c | E | For model evaluation, show a comparison with the development data of the distribution of important predictors (demographics, predictors, and outcome). | p 8 (Results)<br>p 16 (Table 1) |
| **Model development** | 21 | D;E | Specify the number of participants and outcome events in each analysis (e.g., for model development, hyperparameter tuning, model evaluation) | p 8 (Results)<br>p 16 (Table 1) |
| **Model specification** | 22 | D | Provide details of the full prediction model (e.g., formula, code, object, application programming interface) to allow predictions in new individuals and to enable third-party evaluation and implementation, including any restrictions to access or re-use (e.g., freely available, proprietary) | p 6 (Methods: Model architecture) |
| **Model performance** | 23a | D;E | Report model performance estimates with confidence intervals, including for any key subgroups (e.g., sociodemographic). Consider plots to aid presentation. | pp 8–9 (Results)<br>p 16 (Table 2)<br>p 19 (Figure 2) |
| | 23b | D;E | If examined, report results of any heterogeneity in model performance across clusters. See TRIPOD Cluster for additional details. | *Not applicable (not examined)* |

| Section/Topic | Item | Dev/Eval* | Checklist item | Reported on page |
|---|---|---|---|---|
| **Model updating** | 24 | E | Report the results from any model updating, including the updated model and subsequent performance | *Not applicable (no model updating)* |
| **DISCUSSION** | | | | |
| **Interpretation** | 25 | D;E | Give an overall interpretation of the main results, including issues of fairness in the context of the objectives and previous studies | pp 9–11 (Discussion) |
| **Limitations** | 26 | D;E | Discuss any limitations of the study (such as a non-representative sample, sample size, overfitting, missing data) and their effects on any biases, statistical uncertainty, and generalizability | p 11 (Discussion) |
| **Usability of the model in the context of current care** | 27a | D | Describe how poor quality or unavailable input data (e.g., predictor values) should be assessed and handled when implementing the prediction model | p 5 (Methods: Study design and participants)<br>p 5 (Methods: Study design and participants) |
| | 27b | D | Specify whether users will be required to interact in the handling of the input data or use of the model, and what level of expertise is required of users | *Not reported* |
| | 27c | D;E | Discuss any next steps for future research, with a specific view to applicability and generalizability of the model | p 11 (Discussion) |

*D=items relevant only to the development of a prediction model; E=items relating solely to the evaluation of a prediction model; D;E=items applicable to both. TRIPOD-Cluster: Debray et al, BMJ 2023; 380: e071018. AUROC=area under the receiver operating characteristic curve.